\documentclass[10pt, final, journal, letterpaper, twocolumn]{IEEEtran}
\usepackage{amsmath}
\usepackage{amssymb}
\usepackage{amsfonts}
\usepackage{caption}
\usepackage{graphicx}
\usepackage{epsfig}
\usepackage{epsf}
\usepackage{subfigure}
\usepackage{psfrag}
\usepackage{cite}
\usepackage{latexsym}
\usepackage{url}
\usepackage{color}
\usepackage{bm}
\usepackage{cases}
\usepackage{empheq}
\usepackage{graphics}
\usepackage{verbatim}
\usepackage{stfloats}

\usepackage{titlesec}
\usepackage{subcaption}

\usepackage{diagbox}

\begin{document}
\title{UAV-Enabled Secure ISAC Against Dual Eavesdropping Threats: Joint Beamforming\\ and Trajectory Design}
\author{\IEEEauthorblockN{Jianping~Yao,~\IEEEmembership{Member,~IEEE},~Zeyu~Yang,~Zai~Yang,~\IEEEmembership{Senior~Member,~IEEE},~\\Jie~Xu,~\IEEEmembership{Fellow,~IEEE},~and~Tony~Q.~S.~Quek,~\IEEEmembership{Fellow,~IEEE}}
\thanks{J. Yao and Z. Yang are with the School of Information Engineering, Guangdong University of Technology, Guangzhou 510006, China (E-mails: yaojp@gdut.edu.cn, yangzeyugdut@qq.com).}
\thanks{Z. Yang is with the School of Mathematics and Statistics, Xi'an Jiaotong University, Xi'an 710049, China (E-mail: yangzai@xjtu.edu.cn).}
\thanks{J. Xu is with the Future Network of Intelligence Institute (FNii) and the School of Science and Engineering, The Chinese University of Hong Kong, Shenzhen, Shenzhen 518172, China (E-mail: xujie@cuhk.edu.cn). J. Xu is the corresponding author.}
 \thanks{T. Q. S. Quek is with Information Systems Technology and Design, Singapore University of Technology and Design, Singapore 487372, Singapore (E-mail: tonyquek@sutd.edu.sg).}
}
\maketitle

\begin{abstract}
In this work, we study an unmanned aerial vehicle (UAV)-enabled secure integrated sensing and communication (ISAC) system, where a UAV serves as an aerial base station (BS) to simultaneously perform communication with a user and detect a target on the ground, while a dual-functional eavesdropper attempts to intercept the signals for both sensing and communication. Facing the dual eavesdropping threats, we aim to enhance the average achievable secrecy rate for the communication user by jointly designing the UAV trajectory together with the transmit information and sensing beamforming, while satisfying the requirements on sensing performance and sensing security, as well as the UAV power and flight constraints. To address the non-convex nature of the optimization problem, we employ the alternating optimization (AO) strategy, jointly with the successive convex approximation (SCA) and semidefinite relaxation (SDR) methods. Numerical results validate the proposed approach, demonstrating its ability to achieve a high secrecy rate while meeting the required sensing and security constraints.
\end{abstract}


\section{Introduction}
Integrated sensing and communication (ISAC) is regarded as a promising cornerstone technology for future sixth-generation (6G) wireless networks \cite{LiuIntegrated2022}, in which base stations (BSs) can transmit unified ISAC signals and perform ISAC signal processing over shared hardware platforms, thus significantly enhancing the utilization efficiency of spectrum, hardware, and energy resources.
However, in remote regions, such as rural macro areas, or during emergency situations like post-earthquake scenarios or maritime incidents, deploying ground-based BSs can be challenging or even infeasible due to infrastructural limitations and accessibility issues.
With recent advancements in unmanned aerial vehicle (UAV) technology, the exploitation of UAVs has attracted growing interests to provide ISAC services from the sky, in which the UAVs' controllable mobility in the three-dimensional (3D) space is utilized, such that UAVs can approach sensing targets and communication users to enhance the ISAC performance (see e.g., \cite{LyuJoint2022,MengUAV2024,KhaliliEfficient2024,KhaliliEnergy2023}).

Nevertheless, owing to the inherent broadcast characteristics of wireless transmission, the communication and sensing signals in ISAC systems are susceptible to interception, posing significant security challenges for both functionalities. 
To address communication security concerns, physical-layer security (PLS) has been proposed as an effective approach by leveraging the wireless channel properties, which has been extensively studied in prior research \cite{ZhongSecure2018}.
The core idea of these works is to propose an ISAC design to leverage artificial noise (AN) for not only interfering with the eavesdropper but also performing the target sensing, thus enhancing the communication security (see e.g., \cite{WuWhen2023,RenRobust2023,LiuSecure2025}).

On the other hand, there have been only a handful of prior works addressing the sensing security issue in ISAC systems. 
For instance, the work \cite{ZouSecuring2024} considered a communication user acting as a sensing eavesdropper, in which the mutual information (MI) of the authorized sensing receiver is maximized via the joint beamforming design, while ensuring that the MI of the eavesdropper remains below a given threshold.
Furthermore, the paper \cite{RenSecure2024} was the first to jointly consider both PLS and sensing security by focusing on a secure cell-free ISAC system, where several ISAC transmitters collaboratively transmit confidential data to several legitimate communication receivers while conducting target detection under the threat of both communication and sensing eavesdroppers.

Different from prior works that focus on communication secrecy or treat sensing and communication eavesdroppers separately, this work studies a new UAV-enabled secure ISAC system. 
In this system, an aerial dual-functional BS delivers secure data to a legitimate communication receiver while simultaneously performing target sensing, and an eavesdropper attempts to intercept both the communication and sensing signals. 
In contrast to the conventional secure ISAC designs (e.g., \cite{ZouSecuring2024,RenSecure2024}), we exploit the UAV trajectory optimization together with the transmit information and sensing beamforming for enhancing the security performance.
In particular, we maximize the average secrecy rate at the legitimate communication receiver, while ensuring the sensing performance requirements, subject to the UAV power and practical flight constraints, as well as the sensing security constraints.
Notably, while the radar signal serves as undesired interference for the eavesdropper from a communication security perspective, it becomes the desired sensing signal for the eavesdropper that can be exploited for sensing interception. 
This thus introduces a trade-off between communication and sensing security, which distinguishes our model from prior studies that primarily focus on communication secrecy or treat sensing and communication eavesdroppers separately.
To tackle the formulated non-convex problem, we develop an effective approach that leverages a combination of alternating optimization (AO), successive convex approximation (SCA), and semidefinite relaxation (SDR). 
Lastly, we provide numerical results to demonstrate the efficiency of the proposed method.

\section{System Model}
\begin{figure}[t]
 \centering
 \includegraphics[width=0.43\textwidth]{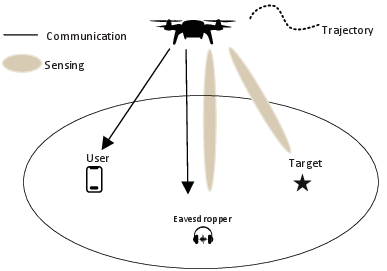} 
 \caption{Illustration of the UAV-enabled secure ISAC system.}
 \label{fig:system_model}
\end{figure}

As shown in Fig. \ref{fig:system_model}, we focus on a fixed-wing UAV-enabled secure ISAC system, where an aerial BS simultaneously transmits data to a legitimate communication user (denoted as $u$) and conducts wireless sensing towards a target (designated as $t$) on the ground, in the presence of a dual-functional eavesdropper (referred to as $e$) intercepting both communication and sensing information.
We assume that the UAV is equipped with a uniform linear array (ULA) consisting of $M$ antennas, arranged in a vertical orientation with respect to the horizontal plane with a constant altitude $D$.
In contrast, all other nodes each have only one antenna.
We consider a 3D Cartesian coordinate system, where the user, the target, and the eavesdropper on the ground with altitude $0$ are fixed at horizontal locations $\boldsymbol s_u = ({x_u},{y_u})$, ${\boldsymbol s_t} = ({x_{t}},{y_{t}})$, and ${\boldsymbol s_e} = ({x_{e}},{y_{e}})$, respectively. 
We assume that the UAV has precise prior knowledge of the positions of both the legitimate ground nodes and the eavesdropper.
\footnote{Note that our proposed framework can be extended to accommodate scenarios where the UAV has only partial or imperfect knowledge of the locations of GRs and/or eavesdroppers. This can be achieved by incorporating robust optimization techniques that account for location uncertainties. Specifically, bounded location error models can be employed, where the actual positions of the nodes are assumed to lie within known uncertainty regions around their estimated locations. Such models have been effectively utilized in prior works to design robust UAV trajectories and transmission strategies that ensure performance guarantees even in the presence of location estimation errors (see, e.g., \cite{ZhongSecure2018}).}
This assumption simplifies the joint design of UAV trajectory and beamforming to offer valuable design insights. 
In practical scenarios, the UAV can acquire the ground nodes' locations through direct reporting from the nodes themselves. 
As for the eavesdropper, its position may be inferred by detecting its signal emissions (if it operates outside the UAV’s network) or obtained from a centralized network controller (when it is part of the same network infrastructure).
\footnote{Even if an eavesdropper operates passively, detection remains feasible since passive receivers inevitably leak minimal radio frequency (RF) energy from their local oscillators \cite{Mukherjee2012}. 
Additionally, the UAVs can leverage the onboard optical cameras and synthetic aperture radar systems to facilitate the detection and localization of potential eavesdroppers through advanced image processing and pattern recognition techniques \cite{ShangUnmanned2019}.
}

We consider a service duration $T$ divided into $N$ equal time slots, each with a duration of ${t_s} = T/N$.
The time slot length is sufficiently short to ensure that the motion states of the UAV remain unchanged within each slot. 
Let $\mathcal N \triangleq \{1, ..., N\}$ denote the collection of slots. 
Therefore, at slot $n \in \mathcal N$, we assume that the UAV's horizontal coordinate is ${\boldsymbol \rho}[n]={(x[n],y[n])}$. 
Let ${{\boldsymbol \rho}_{\rm{I}}} = ({x_{\rm{I}}},{y_{\rm{I}}})$ and ${{\boldsymbol \rho}_{\rm{F}}} = ({x_{\rm{F}}},{y_{\rm{F}}})$ signify the starting and ending horizontal locations of the UAV. 
Let ${v_{\max}}$ and ${V_{\max}} = {v_{\max}}{t_s}$ represent the maximum UAV speed and the maximum displacement within a single time slot, respectively. 
Consequently, we impose the following UAV flight constraints, given as
\begin{equation}
{\boldsymbol \rho}[1] = {{\boldsymbol \rho}_{\rm{I}}},~{\boldsymbol \rho}[N] = {{\boldsymbol \rho}_{\rm{F}}}, \label{eq:initial_final_location}
\end{equation}
\begin{equation}
\left\| {{\boldsymbol \rho}[n + 1] - {\boldsymbol \rho}[n]} \right\| \le {V_{\max }},~\forall n \in \mathcal N. \label{eq:single_time_slot_maximum_displacement}
\end{equation}

Let $a_i[n]$ represent the intended communication signal for the user at slot $n$, ${\boldsymbol b}[n] \in \mathbb{C}^{M \times 1}$ denote the associated transmit beamforming vector, and ${{\boldsymbol a}_s}[n] \in \mathbb{C}^{M \times 1}$ denote the specific wireless sensing signal at slot $n$, which simultaneously serves as AN to interfere with the eavesdropper. 
We assume that the communication signal $a_i[n]$ is independently drawn from a circularly symmetric complex Gaussian (CSCG) distribution. 
Additionally, the wireless sensing signal ${{\boldsymbol a}_s}[n]$ is treated as an independent random vector with a mean of zero and a covariance matrix ${\boldsymbol A}_s[n] = \mathbb{E}({{\boldsymbol a}_s}[n] {{\boldsymbol a}_s^{\rm{H}}}[n]) \succeq \boldsymbol{0}$ \cite{LyuJoint2022}, where $\mathbb{E}\left(\cdot\right)$ denotes the expectation operator and ${{\boldsymbol a}_s^{\rm{H}}}[n]$ represents its conjugate transpose. 

At each slot $n \in \mathcal N$, we denote $\Phi_o({\boldsymbol \rho}[n]), o \in \{u,e,t\}$, as the  steering vector associated with the UAV at location ${\boldsymbol \rho}[n]$ towards ground node ${o}$ as
\begin{align}  
\Phi_o({\boldsymbol \rho}[n]) = {\big[ \phi_{o1}[n],\dots \phi_{om}[n] ,\dots, \phi_{oM}[n] \big]^{\rm{T}}},
\end{align}
where $\phi_{om}[n] = {e^{j2\pi \frac{{k}}{\lambda}(m-1)\cos \theta_o ({\boldsymbol \rho}[n])}}$; 
$\theta_o ({\boldsymbol \rho}[n]) = \text{arccos}{\frac{D}{ \sqrt {  ||{\boldsymbol \rho}[n]-{\boldsymbol s_o}|{|^2}+{D^2}}}}$ denotes the angle of departure (AoD) associated with the location ${\boldsymbol s_o}$;
$\lambda$ refers to the carrier wavelength;
${k} = \lambda/2$ indicates the distance between two adjacent antennas.
Based on the empirical findings reported in \cite{LinThe2018} and the standardized models in Third Generation Partnership Project (3GPP) TR 36.777 \cite{3GPP}, we assume that the UAV operates at a sufficiently high altitude such that the A2G links are predominantly LoS. Accordingly, we adopt a channel model characterized by LoS path loss combined with a directional steering vector.
Then, the channel vector between the UAV and node $o \in \{u,e,t\}$ at slot $n \in \mathcal N$ is given as
\begin{equation}
{\boldsymbol g}_o({\boldsymbol \rho}[n]) = \sqrt {\frac{{{\beta_0}}}{{d_o^2({\boldsymbol \rho}[n])}}} \Phi_o({\boldsymbol \rho}[n]),
\end{equation}
where ${\beta_0}$ represents the channel power gain at a reference distance of 1 meter, 
$d_o({\boldsymbol \rho}[n]) = \sqrt {  ||{\boldsymbol \rho}[n]-{\boldsymbol s_o}|{|^2}+{D^2}}$ is the distance from the UAV to the location ${\boldsymbol s_o}$. 

Accordingly, the received signal at location ${\boldsymbol s_o}, o \in \{u,e,t\}$, at slot $n \in \mathcal N$ is expressed as
\begin{equation}
{z_o}[n] = {\boldsymbol g}_o^{\rm H}({\boldsymbol \rho}[n]) ( {{\boldsymbol b}[n]a_i[n] + }{{\boldsymbol a}_s}[n]) + v_o[n],
\end{equation}
where $v_o[n]$ represents the additive white Gaussian noise (AWGN) at the location ${\boldsymbol s_o}$'s receiver, characterized as a CSCG random variable with a mean of zero and a variance of $\sigma^2$.

The received signal-to-interference-plus-noise ratio (SINR) at the user and the eavesdropper at slot $n$ are respectively given as
\begin{equation}
{\boldsymbol{\gamma}_u}[n] = \frac{|{\boldsymbol g}_u^{\rm{H}}({\boldsymbol \rho}[n]){\boldsymbol b}[n]{|^2}}{{\boldsymbol g}_u^{\rm{H}}({\boldsymbol \rho}[n]){\boldsymbol A}_s[n]{\boldsymbol g}_u({\boldsymbol \rho}[n])+\sigma^2},
\end{equation}
\begin{equation}
{\boldsymbol{\gamma}_e}[n] = \frac{|{\boldsymbol g}_e^{\rm{H}}({\boldsymbol \rho}[n]){\boldsymbol b}[n]{|^2}}{{\boldsymbol g}_e^{\rm{H}}({\boldsymbol \rho}[n]){\boldsymbol A}_s[n]{\boldsymbol g}_e({\boldsymbol \rho}[n]) +\sigma^2}.
\end{equation}

Then, the achievable rates from the UAV to the user and the eavesdropper (in bps/Hz) at slot $n$ are formulated as
\begin{equation}
{R_u}[n] = {\log_2}( {1 + {\boldsymbol{\gamma}_u}[n]} ),
\label{eq:achievable_rate_from_UAV_to_user }
\end{equation}
\begin{equation}
{R_e}[n] = {\log_2}( {1 + {\boldsymbol{\gamma}_e}[n]} ).
\label{eq:achievable_rate_from_UAV_to_eavesdropper}
\end{equation}

Consequently, the secrecy rate from the UAV to the user at slot $n$ is \cite{ZhongSecure2018}
\begin{equation}
{R_s}[n] = [{\log_2}( {1 + {\boldsymbol{\gamma}_u}[n]} )-{\log_2}( {1 + {\boldsymbol{\gamma}_e}[n]} )]^+,
\label{eq:secrecy_rate_from_UAV_to_user}
\end{equation}
where $[u]^+\triangleq \max(u,0)$.

In the considered ISAC system, the UAV intends to detect the target.
To properly illuminate the target, the transmit beampattern gain $\zeta_t[n]$ at the specified sensing location must meet a threshold $\Gamma_{t}$, which is proportional to the square of the distance between the UAV and the target, given as \cite{LyuJoint2022,LiuJoint2020}
\begin{equation}
\begin{array}{l}
\zeta_t[n] = \mathbb{E}[|\Phi_t^{\rm{H}}({\boldsymbol \rho}[n])( {{\boldsymbol b}[n]a_i[n] + }{{\boldsymbol a}_s}[n]){|^2}] \\
= \Phi_t^{\rm{H}}({\boldsymbol \rho}[n])( {{\boldsymbol b}[n]{\boldsymbol b}^{\rm{H}}[n] + {\boldsymbol A}_s[n]} )\Phi_t({\boldsymbol \rho}[n]) \\
\ge \Gamma_{t} {d_t^2}({\boldsymbol \rho}[n]),\forall n \in \mathcal N.
\end{array} 
\end{equation}

Similarly, to ensure sensing security, the transmit beampattern gain $\zeta_e[n]$ at the eavesdropper should not exceed a specific threshold $\Gamma_{e}$, which is proportional to the square of the distance between the UAV and the eavesdropper, given as \cite{LyuJoint2022,LiuJoint2020}
\begin{equation}
\begin{array}{l}
\zeta_e[n]  = \mathbb{E}[|\Phi_e^{\rm{H}}({\boldsymbol \rho}[n])( {{\boldsymbol b}[n]a_i[n] + }{{\boldsymbol a}_s}[n]){|^2}] \\
=\Phi_e^{\rm{H}}({\boldsymbol \rho}[n])( {{\boldsymbol b}[n]{{\boldsymbol b}^{\rm{H}}}[n] + {\boldsymbol A}_s[n]} )\Phi_e({\boldsymbol \rho}[n])\\
 \le \Gamma_{e}{d_e^2}({\boldsymbol \rho}[n]) ,\forall n \in \mathcal N.
\end{array} 
\end{equation}

This study focuses on the joint optimization of the communication beamforming vectors $\{{\boldsymbol b}[n]\} $, the sensing covariance matrices $\{{\boldsymbol A}_s[n]\}$, and the UAV trajectory $\{{\boldsymbol \rho}[n]\} $ to maximize the average secrecy rate, subject to sensing security constraints, sensing constraints, power constraints, and UAV trajectory constraints. 
The problem is formulated as
\begin{subequations} 
\begin{align}
\text{(P1)}:&\mathop {\max }\limits_{\{ {{\boldsymbol b}[n]}, {\boldsymbol A}_s[n],{\boldsymbol \rho}[n]\} } \frac{1}{N}\sum\limits_{n = 1}^N { {R_s}[n] }
\nonumber \\ 
\mathrm{s.t.}~&{\boldsymbol \rho}[1] = {{\boldsymbol \rho}_{\rm{I}}},~{\boldsymbol \rho}[N] = {{\boldsymbol \rho}_{\rm{F}}}, \\
&\left\| {{\boldsymbol \rho}[n + 1] - {\boldsymbol \rho}[n]} \right\| \le {V_{\max }},~\forall n \in \mathcal N,\\
&\zeta_t[n] \ge \Gamma_{t} {d_t^2}({\boldsymbol \rho}[n]),~\forall n \in \mathcal N, \label{eq:sensing_constraint}\\
&\zeta_e[n] \le \Gamma_{e}{d_e^2}({\boldsymbol \rho}[n]) ,\forall n \in \mathcal N, \label{eq:sensing_security_constraint} \\
&{||{\boldsymbol b}[n]|{|^2} + {\rm{tr}}({\boldsymbol A}_s[n])} \le {P_{\max }},\forall n \in \mathcal N, \label{eq:power_constraint} 
\end{align} 
\end{subequations}
where ${P_{\max }}$ represents the UAV's maximum allowable power level.
Since the objective function, along with constraints \eqref{eq:sensing_constraint} and \eqref{eq:sensing_security_constraint} are non-convex, problem (P1) is inherently challenging to be solved directly.

\section{Proposed Solution}
This section presents an effective approach for solving problem (P1) by leveraging convex optimization, SDR, and SCA techniques.

\subsection{Optimization of Information  and Sensing Beamforming}
We first address the optimization of the communication beamforming vectors $\{{\boldsymbol b}[n]\}$ and the sensing covariance matrices $\{{\boldsymbol A}_s[n]\}$ while keeping the UAV trajectory $\{{\boldsymbol \rho}[n]\}$ fixed. 
Under this consideration, problem (P1) is simplified to
\begin{align}
\text{(P2)}:&\mathop {\max }\limits_{\{ {\boldsymbol b}[n],{\boldsymbol A}_s[n]\} } \frac{1}{N}\sum\limits_{n = 1}^N { {R_s}[n] } \nonumber\\
\mathrm{s.t.} &~\eqref{eq:sensing_constraint},~\eqref{eq:sensing_security_constraint}, ~\text{and} ~\eqref{eq:power_constraint}. \nonumber
\end{align}

We define $ {\boldsymbol B}[n] = {\boldsymbol b}[n]{\boldsymbol b}^{\rm H}[n]$ such that rank$({\boldsymbol B}[n]) \le 1$ and ${\boldsymbol B}[n] \succeq \boldsymbol{0}$.
Then problem (P2) is equivalently transformed as
\begin{subequations} 
\begin{align}
\text{(P3)}:&\mathop {\max }\limits_{\{ {\boldsymbol b}[n],{\boldsymbol A}_s[n]\} } \frac{1}{N}\sum\limits_{n = 1}^N {( {R_u}[n] -{R_e}[n])} \nonumber\\
\mathrm{s.t.} 
~&\text{rank}({\boldsymbol B}[n]) \le 1, {\boldsymbol B}[n] \succeq \boldsymbol{0}, \forall n \in \mathcal N,\label{eq:rank_1_constraint}\\
&\Phi_t^{\rm{H}}({\boldsymbol \rho}[n])( {{\boldsymbol B}[n] + {\boldsymbol A}_s[n]} )\Phi_t({\boldsymbol \rho}[n])\nonumber\\
&~~~~~~~~~~~~~~~~~~~~~~~\ge \Gamma_{t} {d_t^2}({\boldsymbol \rho}[n]),\forall n \in \mathcal N,\label{eq:sensing_constraint_P3}\\
&{\Phi_e^{\rm{H}}}({\boldsymbol \rho}[n])( {{\boldsymbol B}[n] + {\boldsymbol A}_s[n]} )\Phi_e({\boldsymbol \rho}[n]) \nonumber\\
&~~~~~~~~~~~~~~~~~~~~~~~\le \Gamma_{e}{d_e^2}({\boldsymbol \rho}[n]) ,\forall n \in \mathcal N,\label{eq:sensing_security_constraint_P3}\\
&{\rm{tr}}({{\boldsymbol B}[n])+ {\rm{tr}}({\boldsymbol A}_s[n])} \le {P_{\max }},\forall n \in \mathcal N.\label{eq:power_constraint_P3} 
\end{align}
\end{subequations}

Problem (P3) remains non-convex due to the non-concave objective function and the rank-one constraint.
To address this, we apply the SDR approach by omitting the rank-one constraint in
\eqref{eq:rank_1_constraint}. 
Subsequently, we handle the non-concave objective function of problem (P3) by applying the SCA technique to achieve a convergent solution iteratively.
At each iteration $l \ge 1$, we derive a lower bound of the objective function under given local point ${\boldsymbol B}^{(l)}[n]$ and ${\boldsymbol A}^{(l)}_s[n]$ through the first-order Taylor expansion, given as

\begin{align}\label{eq:object_function_of_p1}
&{\hat R}^{(l)}[n] \triangleq 
{\log_2}\big({{\rm{tr}}({\boldsymbol g}_u({{\boldsymbol \rho}[n]}){{\boldsymbol g}_u^{\rm{H}}({{\boldsymbol \rho}[n]})}{\boldsymbol B}[n])} \nonumber \\
& + {\rm{tr}}({\boldsymbol g}_u({{\boldsymbol \rho}[n]}){{\boldsymbol g}_u^{\rm{H}}({{\boldsymbol \rho}[n]})}{\boldsymbol A}_s[n]) +\sigma^2\big) \nonumber\\
 &+ {\log_2}\big({\rm{tr}}({\boldsymbol g}_e({{\boldsymbol \rho}[n]}){{\boldsymbol g}_e^{\rm{H}}({{\boldsymbol \rho}[n]})}{\boldsymbol A}_s[n]) +\sigma^2\big)\\
& - \big({\delta}_u^{(l)}[n] 
 + {\rm{tr}}({\Lambda}_u^{(l)}[n]({\boldsymbol A}_s[n] - {\boldsymbol A}_s^{(l)}[n]))\big) \nonumber\\
&{\rm{ - }}\big({\delta}_e^{(l)}[n] + {{\rm{tr}}({\Lambda}_e^{(l)}[n]({\boldsymbol B}[n] - {\boldsymbol B}^{(l)}[n]))}  \nonumber\\
 &+ {\rm{tr}}({\Lambda}_e^{(l)}[n]({\boldsymbol A}_s[n] - {\boldsymbol A}_s^{(l)}[n]))\big), \nonumber
\end{align}
where
\begin{equation}
{\delta}_u^{(l)}[n] \! =\! {\log_2}\big( {\rm{tr}}({{\boldsymbol g}_u({\boldsymbol \rho}[n])}{{\boldsymbol g}_u^{\rm{H}}({\boldsymbol \rho}[n])}{\boldsymbol A}_s^{(l)}[n] ) +\sigma^2 \big),
\label{eq:a_of_p1}
\end{equation}

\begin{equation}
{\Lambda}_u^{(l)}[n] = \frac{{{\log_2}e}~{{\boldsymbol g}_u({\boldsymbol \rho}[n]){{\boldsymbol g}_u^{\rm{H}}({\boldsymbol \rho}[n])}}}{{{\rm{tr}}\big({{\boldsymbol g}_u({\boldsymbol \rho}[n])}{{\boldsymbol g}_u^{\rm{H}}({\boldsymbol \rho}[n])}{\boldsymbol A}_s^{(l)}[n]\big) +\sigma^2}},
\label{eq:B_of_p1}
\end{equation}

\begin{align}
{\delta}_e^{(l)}[n]   &= {\log_2}\big({{\rm{tr}}({{{\boldsymbol g}_e}({{\boldsymbol \rho}[n]})}{{{\boldsymbol g}_e}^{\rm{H}}({{\boldsymbol \rho}[n]})}{\boldsymbol B}^{(l)}[n])} \nonumber \\
&~+{\rm{tr}}({{\boldsymbol g}_e({\boldsymbol \rho}[n])}{{\boldsymbol g}_e^{\rm{H}}({\boldsymbol \rho}[n])}{\boldsymbol A}_s^{(l)}[n] ) +\sigma^2 \big),
\label{eq:ae_of_p1}
\end{align}

\begin{align}
{\Lambda}_e^{(l)}[n] &= {\big({{\log_2}e}~{{\boldsymbol g}_e({\boldsymbol \rho}[n]){{\boldsymbol g}_e^{\rm{H}}({\boldsymbol \rho}[n])}}\big)} \nonumber\\
&~\div \big({{\rm{tr}}({{\boldsymbol g}_e({\boldsymbol \rho}[n])}{{\boldsymbol g}_e^{\rm{H}}({\boldsymbol \rho}[n])}{\boldsymbol B}^{(l)}[n])} \nonumber\\
&~+{{\rm{tr}}({{\boldsymbol g}_e({\boldsymbol \rho}[n])}{{\boldsymbol g}_e^{\rm{H}}({\boldsymbol \rho}[n])}{\boldsymbol A}_s^{(l)}[n]) +\sigma^2}\big).
\label{eq:Be_of_p1}
\end{align}

Consequently, we approximate problem (P3) as a convex problem and solve it iteratively.

\newtheorem{lemma}{Lemma}
\newtheorem{proof}{Proof}
\begin{lemma}\label{lemma1}
With the converged solution ${\boldsymbol B}^*[n]$ and ${\boldsymbol A}^*_s[n]$ obtained by the SCA technique, we can always construct an alternative rank-one solution achieving the same optimal objective, given by
 \begin{equation}\label{w rankone}
{\boldsymbol{\bar b}[n]} = \frac{{\boldsymbol B}^*[n]{\boldsymbol g}_u({\boldsymbol \rho}[n])}{\sqrt{{\boldsymbol g}_u^{\rm{H}}({\boldsymbol \rho}[n]){\boldsymbol B}^*[n]{\boldsymbol g}_u({\boldsymbol \rho}[n])}},
\end{equation}
\begin{equation}\label{W rankone}
{\bar{\boldsymbol{B}}[n]}= {\boldsymbol{\bar b}[n]}{\boldsymbol{\bar b}^{\rm H}[n]},
\end{equation}
\begin{equation}\label{r rankone}
{\bar{\boldsymbol A}}_s[n] = {\boldsymbol B}^*[n] + {\boldsymbol A}^*_s[n] - {\bar{\boldsymbol{B}}}[n],
\end{equation}
where $\text{rank}({\bar{\boldsymbol{B}}}[n])=1$. 

\begin{proof}
See Appendix \ref{appendices_A}.
\end{proof}
\end{lemma}

\subsection{UAV Trajectory Optimization}
With the information and sensing beamforming $\{{\boldsymbol b}[n]\}$ and $\{{\boldsymbol A}_s[n]\}$ fixed, we proceed to optimize the UAV trajectory $\{{\boldsymbol \rho}[n]\}$, for which problem (P1) is reduced to
\begin{subequations}\label{eq:UAV_trajectory_subproblem}
\begin{align}
\text{(P4)}:&\mathop {\max }\limits_{\{ {\boldsymbol \rho}[n]\} }\frac{1}{N}\sum\limits_{n = 1}^N { {{R_s[n]}} }
\nonumber \\
\mathrm{s.t.} \ 
&\Phi_t^{\rm{H}}({\boldsymbol \rho}[n]){{\boldsymbol E}[n]} \Phi_t({\boldsymbol \rho}[n])\ge \Gamma_{t}{d_t^2}({\boldsymbol \rho}[n]), \forall n \in \mathcal N, \label{eq:sensing_constraint_of_P3} \\
&{\Phi_e^{\rm{H}}}({\boldsymbol \rho}[n]){{\boldsymbol E}[n]} \Phi_e({\boldsymbol \rho}[n]) \le \Gamma_{e}{d_e^2}({\boldsymbol \rho}[n]) ,\forall n \in \mathcal N, \label{eq:sensing_security_constraint_of_P3}\\
& ~\eqref{eq:initial_final_location}~\text{and}~\eqref{eq:single_time_slot_maximum_displacement}, \nonumber
\end{align}
\end{subequations}
where ${{\boldsymbol E}[n] = {{\boldsymbol b}[n]{\boldsymbol b}^{\rm{H}}[n] + {\boldsymbol A}_s[n]} }$ is introduced for notational convenience. 
Consequently, we represent the elements in the $i$-th row and $j$-th column of ${\boldsymbol B}[n]$, ${\boldsymbol A}_s[n]$, and ${\boldsymbol E}[n]$ as
$\big[{\boldsymbol B}[n]\big]_{i,j}$, $\big[{\boldsymbol A}_s[n]\big]_{i,j}$, and $\big[{\boldsymbol E}[n]\big]_{i,j}$, 
where their magnitudes are denoted by $\big|\big[{{\boldsymbol B}}[n]\big]_{i,j}\big|$, $\big|\big[{{\boldsymbol{A}}_s}[n]\big]_{i,j}\big|$, and $\big|\big[{\boldsymbol E}[n]\big]_{i,j}\big|$ and their phases are denoted by $\theta_{i,j}^{\boldsymbol B}[n]$, $\theta_{i,j}^{{\boldsymbol{A}}_s}[n]$, and $\theta_{i,j}^{\boldsymbol E}[n]$, respectively.
Problem (P4) is non-convex due to the non-convexity of the objective function in (P4), constraint \eqref{eq:sensing_constraint_of_P3}, and constraint \eqref{eq:sensing_security_constraint_of_P3}. 
To address this, we rewrite the non-concave objective function and constraints as
\begin{align}\label{eq:objective_function_expansion_equation2}
&{R_s}[n]={\log_2}\big( {{\eta_u ({\boldsymbol E}[n],{\boldsymbol \rho}[n])}} \big)+{\log_2}\big( {\xi_e ({\boldsymbol A}_s[n],{\boldsymbol \rho}[n])} 
\big) \nonumber\\
&-{\log_2}\big( {\xi_u ({\boldsymbol A}_s[n],{\boldsymbol \rho}[n])} \big)-{\log_2}\big( {{\eta_e ({\boldsymbol E}[n],{\boldsymbol \rho}[n])}} \big), 
\end{align}
\begin{align}
{\Xi_t ({\boldsymbol E}[n],{\boldsymbol \rho}[n])}/{d_t^2}({\boldsymbol \rho}[n]) \ge \Gamma_{t}, 
\label{eq:sensing_constraint_rewrite_again}
\end{align}
\begin{align}
{\Xi_e ({\boldsymbol E}[n],{\boldsymbol \rho}[n])}/{d_e^2}({\boldsymbol \rho}[n]) \le \Gamma_{e}, 
\label{eq:sensing_security_constraint_rewrite_again}
\end{align}
where $\Xi_o{ ({\boldsymbol E}[n],{\boldsymbol \rho}[n])} = {\eta_o ({\boldsymbol E}[n],{\boldsymbol \rho}[n])} -\frac{\sigma^2}{{\beta_0} }{d_o^2}({\boldsymbol \rho}[n]), o \in \{u,e,t\}$;
${\eta_u ({\boldsymbol E}[n],{\boldsymbol \rho}[n])}$, $\xi_u ({\boldsymbol A}_s[n],{\boldsymbol \rho}[n])$, ${\eta_e ({\boldsymbol E}[n],{\boldsymbol \rho}[n])}$, $\xi_e ({\boldsymbol A}_s[n],{\boldsymbol \rho}[n])$, and ${\eta_t ({\boldsymbol E}[n],{\boldsymbol \rho}[n])}$ are given as 
\begin{align}
\nonumber\eta_o ({\boldsymbol E}[n]&,{\boldsymbol \rho}[n]) = \sum\limits_{p = 1}^M {{[{\boldsymbol E}[n]]}_{p ,p }} \\
\nonumber&+ 2\sum\limits_{i = 1}^M {\sum\limits_{j = i + 1}^M {|{{[{\boldsymbol E}[n]]}_{i,j}}|} } \cos \big(\theta_{i,j}^{\boldsymbol E}[n] \!+\! \frac{\pi (j- i)D}{d_o({{\boldsymbol \rho}[n]})}\big)
\\&+\frac{\sigma^2}{{\beta_0} }{d_o^2}({\boldsymbol \rho}[n]), o \in \{u,e,t\}.
\label{eq:eta_of_c}
\end{align}
\begin{align}
\nonumber\xi_o ({\boldsymbol A}_s[n]&,{\boldsymbol \rho}[n]) = \sum\limits_{p = 1}^M {{[{\boldsymbol A}_s[n]]}_{p ,p }} 
\\\nonumber&+  2\sum\limits_{i = 1}^M {\sum\limits_{j = i + 1}^M {|{{[{\boldsymbol A}_s[n]]}_{i,j}}|} } \cos \big(\theta_{i,j}^{{\boldsymbol A}_s}[n] + \frac{\pi (j- i)D}{d_o({{\boldsymbol \rho}[n]})}\big)
\\&+\frac{\sigma^2}{{\beta_0} }{d_o^2}({\boldsymbol \rho}[n]), o \in \{u,e,t\}.
\label{eq:mu_of_c}
\end{align}

Next, we introduce the trust-region-based SCA method, which is executed iteratively. 
Considering a specific iteration $l$ with a local trajectory point ${{\boldsymbol \rho}^{(l)}}[n]$, we approximate \eqref{eq:objective_function_expansion_equation2}, \eqref{eq:sensing_constraint_rewrite_again}, and \eqref{eq:sensing_security_constraint_rewrite_again} applying the first-order Taylor expansion as
\begin{align}
&\bar R_s^{(l)}[n] \buildrel \Delta \over ={\log_2}\big( {\eta_u ({\boldsymbol E}[n],{{\boldsymbol \rho}^{(l)}[n]})} \big) - {\log_2}\big( { \xi_u ({\boldsymbol A}_s[n],{{\boldsymbol \rho}^{(l)}[n]}) } \big) \nonumber\\
&-\big ({\log_2}( {\eta_e ({\boldsymbol E}[n],{{\boldsymbol \rho}^{(l)}[n]})} ) - {\log_2}( { \xi_e ({\boldsymbol A}_s[n],{ {\boldsymbol \rho}^{(l)}[n]}) } )\big) \nonumber\\
& + ({\boldsymbol \varrho}_u^{{{(l)}^{\rm{{H}}}}}[n] - {\boldsymbol \varrho}_e^{{{(l)}^{\rm{{H}}}}}[n])({\boldsymbol \rho}[n] - {\boldsymbol \rho}^{(l)}[n]), \label{eq:equsetion2_First-order_Taylor_expansion}
\end{align}
\begin{equation}
\begin{array}{l}
\frac{{\Xi_t ({\boldsymbol E}[n],{\boldsymbol \rho}^{(l)}[n])}}{{ {d_t^2}({\boldsymbol \rho}^{(l)}[n])}} + 
\frac{{{\tau}_t^{(l)}[n]{ {d_t^2}({\boldsymbol \rho}^{(l)}[n])}- 2{\Xi_t ({\boldsymbol E}[n],{\boldsymbol \rho}^{(l)}[n])}({\boldsymbol \rho}^{(l)}[n] - {{\boldsymbol s_t}})}}{{ {d_t^4}({\boldsymbol \rho}^{(l)}[n])}}\\ 
\times({\boldsymbol \rho}[n] - {\boldsymbol \rho}^{(l)}[n]) \ge \Gamma_{t}, 
\label{eq:sensing_constraint_convex}
\end{array}
\end{equation}
\begin{equation}
\begin{array}{l}
\frac{{\Xi_e ({\boldsymbol E}[n],{\boldsymbol \rho}^{(l)}[n])}}{{ {d_e^2}({\boldsymbol \rho}^{(l)}[n])}} + 
\frac{{{\tau}_e^{(l)}[n]{ {d_e^2}({\boldsymbol \rho}^{(l)}[n])} - 2{\Xi_e ({\boldsymbol E}[n],{\boldsymbol \rho}^{(l)}[n])}({{\boldsymbol \rho}^{(l)}}[n] - {{\boldsymbol s_e}})}}{{ {d_e^4}({\boldsymbol \rho}^{(l)}[n])}}\\
\times({\boldsymbol \rho}[n] - {{\boldsymbol \rho}^{(l)}}[n])
 \le \Gamma_{e}, 
\label{eq:sensing_security_constraint_convex}
\end{array}
\end{equation}
where ${\boldsymbol \varrho}_u^{(l)}[n]$, ${\boldsymbol \varrho}_e^{(l)}[n]$, $\iota_o ({\boldsymbol E}[n], {{\boldsymbol \rho}^{(l)}[n]})$, $\varsigma_o ({\boldsymbol A}_s[n],{{\boldsymbol \rho}^{(l)}[n]})$, and ${\tau}_o^{(l)}[n]$ are written as
\begin{align}
\nonumber{\boldsymbol \varrho}_o^{(l)}[n] =& \frac{{{{\log }_2}e}}{{\eta_o ({\boldsymbol E}[n],{{\boldsymbol \rho}^{(l)}[n]})} } {\iota_o ({\boldsymbol E}[n],{ {\boldsymbol \rho}^{(l)}[n]}) ({\boldsymbol \rho}^{(l)}[n] - {{\boldsymbol s_o}})} 
 \\&\nonumber\!-\! \frac{{{{\log }_2}e}}{{ \xi_o ({\boldsymbol A}_s[n],{{\boldsymbol \rho}^{(l)}[n]})}} {\varsigma_o ({\boldsymbol A}_s[n],{ {\boldsymbol \rho}^{(l)}[n]}) }
 \\&\times({\boldsymbol \rho}^{(l)}[n] - {{\boldsymbol s_o}}), o \in \{u,e,t\}.
\label{eq:de_second_term_of_equation2}
\end{align}
\begin{align}\label{eq:first-derivative_of_eta}
\iota_o ({\boldsymbol E}&[n],{{\boldsymbol \rho}^{(l)}[n]})
 = \frac{2\sigma^2}{{\beta_0}}+\sum\limits_{i = 1}^M {\sum\limits_{j = i + 1}^M {2\pi |{{[{\boldsymbol E}[n]]}_{i,j}}|}} 
\\\nonumber&\times \sin \big(\theta_{i,j}^{\boldsymbol E}[n] + \frac{\pi(j- i)D}{{ {d_o}({\boldsymbol \rho}^{(l)}[n])}}\big)
 \frac{(j- i)D}{{ {d_o^3}({\boldsymbol \rho}^{(l)}[n])}}, o \in \{u,e,t\}. 
\end{align}
\begin{align}\label{eq:first-derivative_of_xi}
\varsigma_o ({\boldsymbol A}_s&[n],{{\boldsymbol \rho}^{(l)}[n]})
 =  \frac{2\sigma^2}{{\beta_0}}+\sum\limits_{i = 1}^M {\sum\limits_{j = i + 1}^M {2\pi |{{[{\boldsymbol A}_s[n]]}_{i,j}}|}}
 \\\nonumber&\times\sin \big(\theta_{i,j}^{{\boldsymbol A}_s}[n] +\frac{ \pi(j- i)D}{{ {d_o}({\boldsymbol \rho}^{(l)}[n])}}\big)\frac{{(j- i)D}}{{ {d_o^3}({\boldsymbol \rho}^{(l)}[n])}} , o \in \{u,e,t\}. 
\end{align}
\begin{align}
\nonumber{\tau}_o^{(l)}[n] =&
 \sum\limits_{i = 1}^M {\sum\limits_{j = i + 1}^M {2\pi|{{[{\boldsymbol E}[n]]}_{i,j}}|\sin \big(\theta_{i,j}^{\boldsymbol E}[n] + \frac{\pi (j- i)D}{{d_o}({\boldsymbol \rho}^{(l)}[n])}\big)} } 
\\&\times\frac{{(j- i)D}}{{ {d_o^3}({\boldsymbol \rho}^{(l)}[n])}}({\boldsymbol \rho}^{(l)}[n] - {{\boldsymbol s_o}}), o \in \{u,e,t\}.
\label{eq:first-derivative_of_first_part_of_eta}
\end{align}

To maintain the accuracy of the approximation, we introduce a set of trust region constraints as
\begin{equation}
||{\boldsymbol \rho}^{(l)}[n] - {\boldsymbol \rho}^{(l-1)}[n]|| \le {\psi^{(l)}},\forall n \in \mathcal N,
\label{eq:trust_region_radius}
\end{equation}
where ${\psi^{(l)}}$ represents the trust region's radius. 
Notably, theoretically, reduce the radius ${\psi^{(l)}}$ to a sufficiently small value, which guarantees the convergence of the iterative procedure.

Ultimately, by substituting the non-concave objective function of problem (P4) and the non-convex constraints \eqref{eq:sensing_constraint_of_P3} and \eqref{eq:sensing_security_constraint_of_P3} with their respective approximate forms as given in \eqref{eq:equsetion2_First-order_Taylor_expansion}, \eqref{eq:sensing_constraint_convex}, and \eqref{eq:sensing_security_constraint_convex}, and incorporating the trust region constraints in \eqref{eq:trust_region_radius}, we derive the convex approximation of problem (P4) in the $l$-th iteration as follows, which can be efficiently solved using CVX.
\begin{align}
\text{(P5.l)}:&\mathop {\max }\limits_{\{ {\boldsymbol \rho}[n]\} } \frac{1}{N}\sum\limits_{n = 1}^N { {\bar R_s^{(l)}[n]} } \\
\mathrm{s.t.}&~\eqref{eq:initial_final_location},~\eqref{eq:single_time_slot_maximum_displacement},~\eqref{eq:sensing_constraint_convex},~\eqref{eq:sensing_security_constraint_convex},~\eqref{eq:trust_region_radius}.
\nonumber
\end{align}

To sum up, we solve for the communication beamforming vectors $\{{\boldsymbol b}[n] \}$, the sensing covariance matrices $\{{\boldsymbol A}_s[n]\} $ and the UAV trajectory $\{{\boldsymbol \rho}[n]\}$ in an alternating manner.
Since both subproblems can be guaranteed to converge,  we finally obtain an efficient solution to problem (P1).

\section{Numerical Results}\label{simulation_section}
In this section, we present numerical results to assess the effectiveness of the proposed algorithm. 
Unless stated otherwise, the simulation settings are as follows:
${{\boldsymbol \rho}_I} =[300,400,200]^{\rm{T}}~\text{m}$, ${{\boldsymbol \rho}_F} =[300,600,200]^{\rm{T}}~\text{m}$, ${{\boldsymbol s_t}} = [250,480,0]^{\rm{T}}~\text{m}$, 
${{\boldsymbol s_u}} = [250,520,0]^{\rm{T}}~\text{m}$, 
${{\boldsymbol s_e}} = [350,500,0]^{\rm{T}}~\text{m}$, 
$T = 12~{\rm{s}}$,
${t_s} = 0.5~{\rm{s}}$,
$N = 24$,
${v_{\max}} = 25~{\rm{m/s}}$,
$M = 4$, $\Gamma_{t}=\Gamma_{e}=10^{-6}$,
${P_{\max }} = 1~{\rm{W}}$, 
$\beta_0 {\rm{ = - 30~\rm{dBm}}}$,
and ${\sigma^2} = - 90~{\rm{dBm}}$.
To facilitate comparison, we evaluate three baseline approaches as follows.
\begin{itemize}
\item {\bf Straight-flight trajectory with beamforming optimization}: 
The UAV adopts the straight-flight trajectory, traveling at a uniform speed $||{{\boldsymbol \rho}_I}-{{\boldsymbol \rho}_F}||/T$ from the starting position to the destination. 
Based on the straight-flight trajectory, the UAV dynamically adjusts the transmit information and sensing beamforming by solving problem (P2).

\item {\bf Trajectory design with maximum ratio transmission (MRT) beamforming}: 
The UAV optimizes the trajectory by solving problem (P4) by considering the simple MRT information beamforming with ${\boldsymbol b}[n] = \sqrt{\min{({P_{\max }}},{P_c})}{\boldsymbol g}_u({\boldsymbol \rho}[n])/||{\boldsymbol g}_u({\boldsymbol \rho}[n])||,~\forall n \in \mathcal N$, in which $P_c$ is the maximum transmission power that satisfies the sensing security threshold at the eavesdropper.

\item {\bf Benchmark without sensing security}: The UAV jointly optimizes its trajectory and the beamforming in problem (P1) by ignoring the sensing security constraint \eqref{eq:sensing_security_constraint}.

\end{itemize}

\begin{figure}[!h] 
 \centering
 \includegraphics[width=0.43\textwidth]{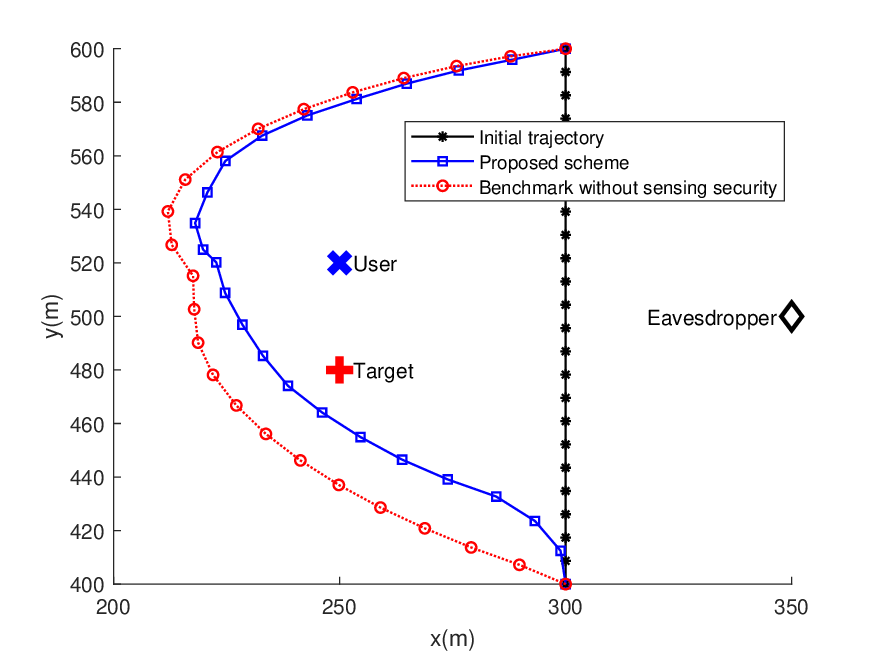} 
 \caption{Trajectories obtained by our proposed design and the benchmark without sensing security.}
 \label{fig:optimized_trajectory}
\end{figure}
\begin{figure}[!h] 
 \centering
 \includegraphics[width=0.43\textwidth]{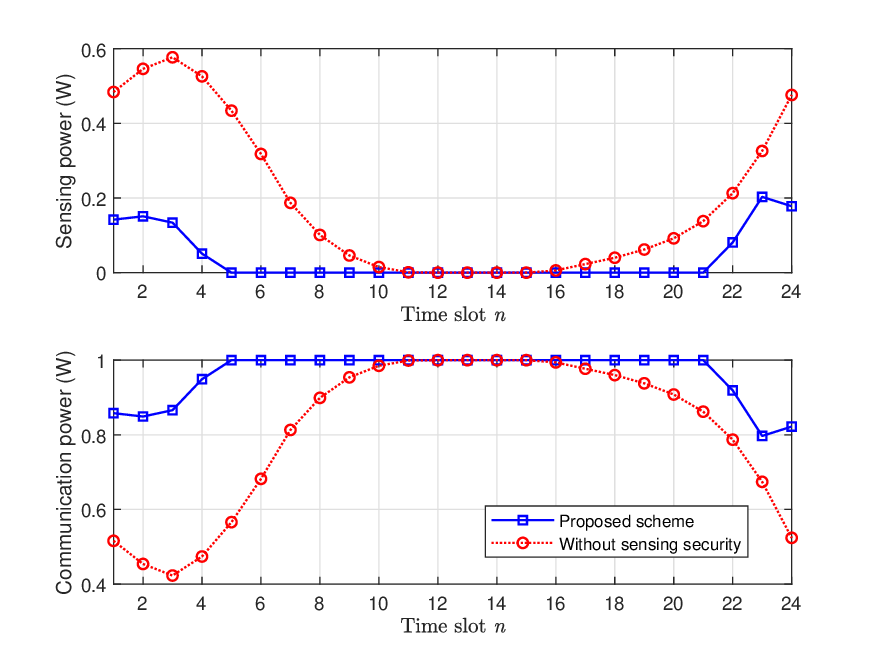} 
 \caption{Power allocation between sensing and communication signals over time by our proposed design and the benchmark without sensing security.}
 \label{fig:signal_power_versus_time_slot}
\end{figure}

Fig. \ref{fig:optimized_trajectory} compares the trajectory obtained by our proposed design and the benchmark without sensing security. 
In both schemes, the UAV is observed to follow arc-like paths that move towards the legitimate user and target while avoiding the eavesdropper, in order to prevent information and sensing leakage.
It is also observed that the trajectory obtained by the proposed design is closer to the legitimate nodes.
This is because when sensing security is considered, most of the power is concentrated in the communication signal and reused for target sensing as shown in Fig. \ref{fig:signal_power_versus_time_slot}. 
Compared to the benchmark without sensing security, the beam is not as focused, so the UAV needs to be closer to the two legitimate nodes in order to cover both.

\begin{figure}[!h]
    \centering 
    \subfigure[Proposed scheme]{\includegraphics[width=0.43\textwidth]{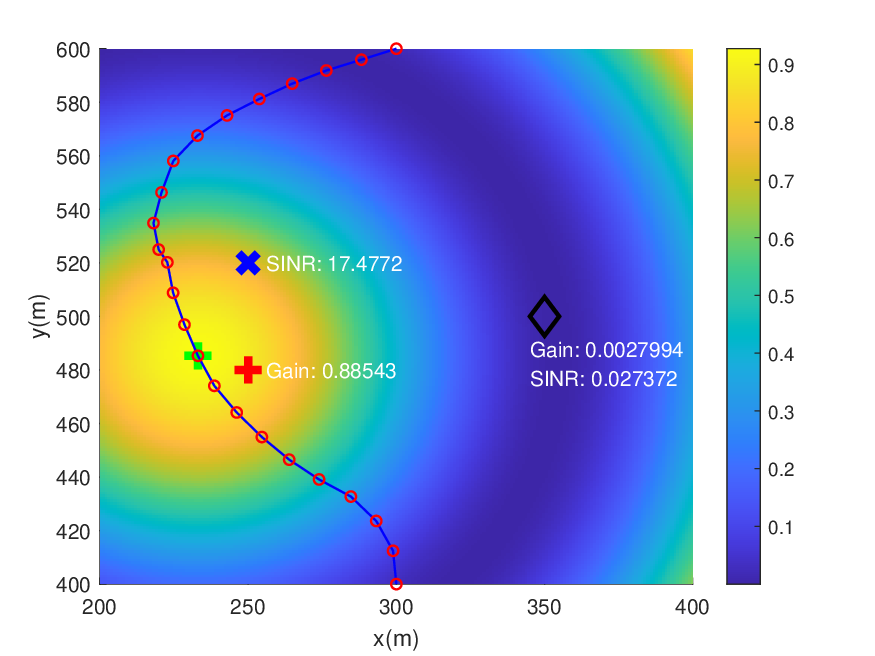}}
    \subfigure[Benchmark without sensing security]{\includegraphics[width=0.43\textwidth]{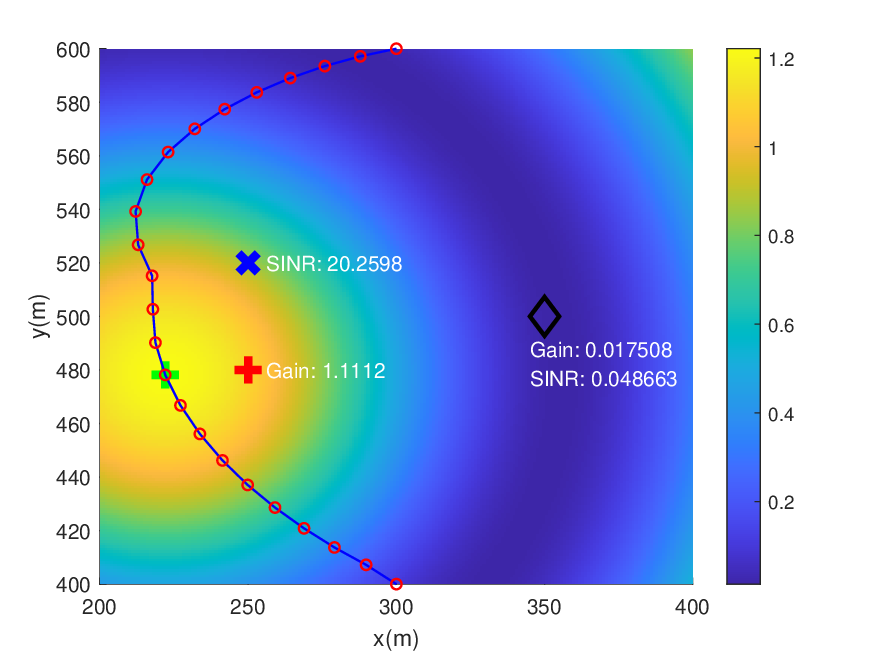}}
    \caption{Achieved beampattern gain at time slot $10$.}
    \label{fig:achieved_beampattern_gain_heatmap}
\end{figure}
Fig. \ref{fig:achieved_beampattern_gain_heatmap} presents the beampattern gain at time slot $10$ in space by our proposed design and the benchmark without sensing security. 
It is observed that both the user and target are within the high-gain region, while the eavesdropper is located in the low-gain region, thus ensuring communication and sensing security.
It is also observed that by comparing the two sub-figures, the SINR and beampattern gain of the user and eavesdropper under the benchmark without sensing security are higher than those of the proposed design.
When sensing security is not considered, the primary objective focuses on enhancing the disparity in the rate between the legitimate and eavesdropping channels to enhance the secrecy rate. 
This results in a higher secrecy rate but also introduces potential sensing security vulnerabilities, which aligns with the findings in Fig. \ref{fig:optimized_trajectory}.

\begin{figure}[!h] 
 \centering
 \includegraphics[width=0.43\textwidth]{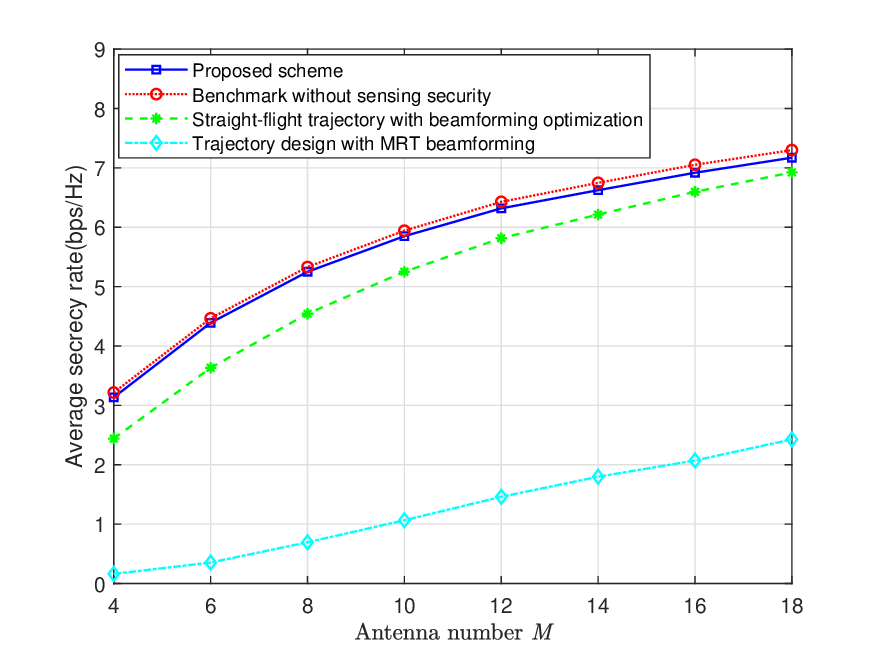} 
 \caption{Average secrecy rate versus the antenna number $M$ over 100 Monte Carlo runs, where the locations of the eavesdropper and ground nodes are randomly generated within the desirable region $[200,400]\times[400,600]$.}
 \label{fig:rate_versus_Number_of_antennas}
\end{figure}
Fig. \ref{fig:rate_versus_Number_of_antennas} illustrates the relationship between the antenna number $M$ and the average secrecy rate.
It is observed that the average secrecy rate achieved by all four schemes increases as the antenna number grows, since additional antennas provide more degrees of freedom and array gains.
By directing energy more accurately toward intended receivers and minimizing leakage to potential eavesdroppers, the system's overall security performance is significantly bolstered.
It is also observed that our proposed scheme significantly outperforms the straight-flight trajectory with beamforming optimization scheme and trajectory design with MRT beamforming scheme.
This superiority stems from the synergistic integration of trajectory planning and beamforming design, which allows the UAV to adapt its path and signal transmission dynamically in response to environmental conditions and potential threats. 
By jointly optimizing these parameters, the system can maintain stronger legitimate links while effectively mitigating the risk of eavesdropping, leading to enhanced secrecy rates compared to schemes that optimize these aspects in isolation.
Additionally, the observed disparity in secrecy rate between our proposed design and the benchmark without sensing security highlights the trade-off between achieving high secrecy rates and ensuring robust protection against sensing eavesdropping.
Incorporating sensing security measures necessitates additional resource allocation and system constraints, which can limit the maximum achievable secrecy rate. However, this trade-off is essential to safeguard sensitive information against adversaries attempting to exploit sensing mechanisms for eavesdropping.

This work considered a UAV-enabled secure ISAC system, where an aerial dual-functional BS simultaneously performs secure communication with a communication user and performs radar sensing of a target, in the presence of an eavesdropper intercepting both information and sensing. 
We focused on maximizing the average achievable secrecy rate through the optimization of the UAV trajectory, as well as the transmit information and sensing beamforming.
To extend our approach to other configurations (e.g., with multiple users over complex dynamic environments) presents promising avenues for future research. This includes developing real-time adaptive algorithms to handle dynamic environments, incorporating robust optimization techniques to address uncertainties in CSI, and accounting for realistic UAV mobility and energy constraints. Furthermore, exploring the integration of multiple antennas and pursuing global optimality solutions could enhance system performance. These directions aim to improve the practical applicability and theoretical robustness of UAV-enabled secure ISAC systems.

\appendices
\section{Proof of Lemma \ref{lemma1}}\label{appendices_A}
With the obtained solution ${\boldsymbol B}^*[n]$  and ${\boldsymbol A}_s^*[n]$ to problem (P3),  we can construct the following solutions, shown as
\begin{align}\label{w rankone}
{\boldsymbol{\bar b}[n]} = \frac{{\boldsymbol B}^*[n]{{\boldsymbol g}_u({{\boldsymbol \rho}[n]})}}{\sqrt{{{\boldsymbol g}_u^{\rm{H}}({{\boldsymbol \rho}[n]})}{\boldsymbol B}^*[n]{{\boldsymbol g}_u({{\boldsymbol \rho}[n]})}}},
\end{align}
\begin{align}\label{W rankone}
{\bar{\boldsymbol{B}}[n]}= {\boldsymbol{\bar b}[n]}{\boldsymbol{\bar b}^{\rm H}[n]},
\end{align}
\begin{align}\label{r rankone}
{\bar{\boldsymbol A}}_s[n] = {\boldsymbol B}^*[n]  + {{\boldsymbol A}_s^*[n]} - {\bar{\boldsymbol{B}}[n]}.
\end{align}

In the following, we will prove that the new constructed solution ${\bar{\boldsymbol{B}}[n]}$ and ${\bar{\boldsymbol A}}_s[n]$ is feasible for problem (P3) with the same objective value as that of the solution ${\boldsymbol B}^*[n]$  and ${\boldsymbol A}_s^*[n]$.
We can easily get that ${\bar{\boldsymbol{B}}[n]}$ is positive semidefinite and rank-one. 
Next, we will show that ${\bar{\boldsymbol A}}_s[n]$ is also positive semidefinite.

For any ${\boldsymbol w} \in \mathbb{C}^{M \times 1}$, it holds that
\begin{align}
&\nonumber{{\boldsymbol{w}}^{\rm{H}}}\left( {{\boldsymbol B}^*[n] - {\bar{\boldsymbol{B}}[n]}} \right){\boldsymbol{w}}\\
\nonumber =& {{\boldsymbol{w}}^{\rm{H}}}{\boldsymbol B}^*[n]{\boldsymbol{w}} - {\left| {{{\boldsymbol{w}}^{\rm{H}}}{\boldsymbol B}^*[n]{{\boldsymbol g}_u^{\rm{H}}({{\boldsymbol \rho}[n]})}} \right|^2}\\
 &{\left( {{{\boldsymbol g}_u^{\rm{H}}({{\boldsymbol \rho}[n]})}{\boldsymbol B}^*[n]{{\boldsymbol g}_u({{\boldsymbol \rho}[n]})}} \right)^{ - 1}}.
\end{align}

According to the Cauchy-Schwarz inequality, we have

\begin{align}
&\nonumber{\left| {{{\boldsymbol{w}}^{\rm{H}}}{\boldsymbol B}^*[n]{{\boldsymbol g}_u({{\boldsymbol \rho}[n]})}} \right|^2}{\left( {{{\boldsymbol g}_u^{\rm{H}}({{\boldsymbol \rho}[n]})}{\boldsymbol B}^*[n]{{\boldsymbol g}_u({{\boldsymbol \rho}[n]})}} \right)^{ - 1}}\\
\nonumber =& {\left| {{{\boldsymbol{w}}^{\rm{H}}}{\boldsymbol b}^*[n]{\boldsymbol b}^*{^{\rm H}}[n]{{\boldsymbol g}_u({{\boldsymbol \rho}[n]})}} \right|^2}{\left( {{{\boldsymbol g}_u^{\rm{H}}({{\boldsymbol \rho}[n]})}{\boldsymbol B}^*[n]{{\boldsymbol g}_u({{\boldsymbol \rho}[n]})}} \right)^{ - 1}}\\
\nonumber \le& {\left| {{{\boldsymbol{w}}^{\rm{H}}}{\boldsymbol b}^*[n]} \right|^2}{\left| {{{\boldsymbol g}_u^{\rm{H}}({{\boldsymbol \rho}[n]})}{\boldsymbol b}^*[n]} \right|^2}{\left( {{{\boldsymbol g}_u^{\rm{H}}({{\boldsymbol \rho}[n]})}{\boldsymbol B}^*[n]{{\boldsymbol g}_u({{\boldsymbol \rho}[n]})}} \right)^{ - 1}}\\
\nonumber =& {{\boldsymbol{w}}^{\rm{H}}}{\boldsymbol b}^*[n]{\boldsymbol b}^*{^{\rm H}}[n]{\boldsymbol{w}}\\
 =& {{\boldsymbol{w}}^{\rm{H}}}{\boldsymbol B}^*[n]{\boldsymbol{w}}.
\end{align}

Thus, we have
\begin{align} \label{W pos-semi}
{{\boldsymbol{w}}^{\rm{H}}}({\boldsymbol B}^*[n] - {\bar{\boldsymbol{B}}[n]}){\boldsymbol{w}} \ge 0.
\end{align}

According to \eqref{W pos-semi}, we have ${\boldsymbol B}^*[n] - {\bar{\boldsymbol{B}}[n]}\succeq 0$. Based on this fact together with ${\bar{\boldsymbol A}}_s^*[n]\succeq 0$, it follows that ${\bar{\boldsymbol A}}_s[n] = {\boldsymbol B}^*[n] + {\boldsymbol A}_s^*[n] - {\bar {\boldsymbol B}[n]}$ should be positive semidefinite.

In addition, we need to prove that substituting the reconstructed solution $\bar{\boldsymbol{B}}[n]$ and $\bar{\boldsymbol A}_s[n]$ into the original problem still yields the same objective function value satisfying the corresponding constraints \eqref{eq:sensing_constraint_P3}, \eqref{eq:sensing_security_constraint_P3}, and \eqref{eq:power_constraint_P3}.
The proof is shown as follows.
\begin{align}
&\nonumber\Phi_t^{\rm{H}}({\boldsymbol \rho}[n])( {{\bar{\boldsymbol{B}}[n]} + {\bar{\boldsymbol A}}_s[n]} )\Phi_t({\boldsymbol \rho}[n]) \\
\nonumber=&\Phi_t^{\rm{H}}({\boldsymbol \rho}[n])( {{\bar{\boldsymbol{B}}[n]} + {\boldsymbol B}^*[n]  + {{\boldsymbol A}_s^*[n]} - {\bar{\boldsymbol{B}}[n]}} )\Phi_t({\boldsymbol \rho}[n])\\
\nonumber=&\Phi_t^{\rm{H}}({\boldsymbol \rho}[n])( {{\boldsymbol B}^*[n]  + {{\boldsymbol A}_s^*[n]}} )\Phi_t({\boldsymbol \rho}[n])\\
\ge& \Gamma_{t} {d_t^2}({\boldsymbol \rho}[n]),
\end{align}

\begin{align}
\nonumber&\Phi_e^{\rm{H}}({\boldsymbol \rho}[n])( {{\bar{\boldsymbol{B}}[n]} + {\bar{\boldsymbol A}}_s[n]} )\Phi_e({\boldsymbol \rho}[n]) \\
\nonumber=&\Phi_e^{\rm{H}}({\boldsymbol \rho}[n])( {{\bar{\boldsymbol{B}}[n]} + {\boldsymbol B}^*[n]  + {{\boldsymbol A}_s^*[n]} - {\bar{\boldsymbol{B}}[n]}} )\Phi_e({\boldsymbol \rho}[n])\\
\nonumber=&\Phi_e^{\rm{H}}({\boldsymbol \rho}[n])( {{\boldsymbol B}^*[n]  + {{\boldsymbol A}_s^*[n]}} )\Phi_e({\boldsymbol \rho}[n])\\
\le& \Gamma_{e}{d_e^2}({\boldsymbol \rho}[n]),
\end{align}
\begin{align}
\nonumber&{\rm{tr}}({{\bar{\boldsymbol{B}}[n]})+ {\rm{tr}}({\bar{\boldsymbol A}}_s[n])}\\ 
\nonumber=& {\rm{tr}}({\bar{\boldsymbol{B}}[n]} + \bar{\boldsymbol A}_s[n] ) \\
\nonumber=& {\rm{tr}}({\bar{\boldsymbol{B}}[n]} + {\boldsymbol B}^*[n]  + {{\boldsymbol A}_s^*[n]} - {\bar{\boldsymbol{B}}[n]}) \\
\nonumber=& {\rm{tr}}({\boldsymbol B}^*[n]  + {{\boldsymbol A}_s^*[n]})\\
\nonumber=& {\rm{tr}}({\boldsymbol B}^*[n]) + {\rm{tr}}({{\boldsymbol A}_s^*[n]})\\
\le& {P_{\max }}.
\end{align}

For the proof of the objective function, we can first derive the following equations
\begin{align}
\nonumber&{{\boldsymbol g}_u^{\rm{H}}({{\boldsymbol \rho}[n]})}{\bar{\boldsymbol{B}}[n]}{{\boldsymbol g}_u({{\boldsymbol \rho}[n]})} \\
\nonumber=&{{\boldsymbol g}_u^{\rm{H}}({{\boldsymbol \rho}[n]})} {\boldsymbol{\bar b}[n]}{\boldsymbol{\bar b}^{\rm H}[n]}{{\boldsymbol g}_u({{\boldsymbol \rho}[n]})} \\
=&{{\boldsymbol g}_u^{\rm{H}}({{\boldsymbol \rho}[n]})}{{\boldsymbol{B}^*}[n]}{{\boldsymbol g}_u({{\boldsymbol \rho}[n]})},
\end{align}

\begin{align}
\nonumber&{{\boldsymbol g}_e^{\rm{H}}({{\boldsymbol \rho}[n]})}{\bar{\boldsymbol{B}}[n]}{{\boldsymbol g}_e({{\boldsymbol \rho}[n]})} \\
\nonumber=&{{\boldsymbol g}_e^{\rm{H}}({{\boldsymbol \rho}[n]})} {\boldsymbol{\bar b}[n]}{\boldsymbol{\bar b}^{\rm H}[n]}{{\boldsymbol g}_e({{\boldsymbol \rho}[n]})} \\
=&{{\boldsymbol g}_e^{\rm{H}}({{\boldsymbol \rho}[n]})}{{\boldsymbol{B}^*}[n]}{{\boldsymbol g}_e({{\boldsymbol \rho}[n]})}.
\end{align}

For the first two terms of Eq. \eqref{eq:object_function_of_p1}, it can be rewritten as
\begin{align}\label{first_two_terms_object_function}
\nonumber&{\log_2}\big({{\rm{tr}}({{\boldsymbol g}_u({{\boldsymbol \rho}[n]})}{{\boldsymbol g}_u^{\rm{H}}({{\boldsymbol \rho}[n]})}{\bar{\boldsymbol{B}}[n]}}) \\
 \nonumber+& {\rm{tr}}({{\boldsymbol g}_u({\boldsymbol \rho}[n])}{{\boldsymbol g}_u^{\rm{H}}({\boldsymbol \rho}[n])}{\bar{\boldsymbol A}}_s[n]) +\sigma^2\big)\\
 \nonumber+& {\log_2}\big({\rm{tr}}({{\boldsymbol g}_e({\boldsymbol \rho}[n])}{{\boldsymbol g}_e^{\rm{H}}({\boldsymbol \rho}[n])}{\bar{\boldsymbol A}}_s[n]) +\sigma^2\big)\\
 \nonumber=&{\log_2}\big({{\rm{tr}}({{\boldsymbol g}_u({{\boldsymbol \rho}[n]})}{{\boldsymbol g}_u^{\rm{H}}({{\boldsymbol \rho}[n]})}{\bar{\boldsymbol{B}}[n]}}) \\
 \nonumber+& {\rm{tr}}({{\boldsymbol g}_u({\boldsymbol \rho}[n])}{{\boldsymbol g}_u^{\rm{H}}({\boldsymbol \rho}[n])}
({\boldsymbol B}^*[n]  + {\boldsymbol A}_s^*[n] - {\bar{\boldsymbol{B}}[n]})) +\sigma^2\big)\\
 \nonumber+& {\log_2}\big({\rm{tr}}({{\boldsymbol g}_e({\boldsymbol \rho}[n])}{{\boldsymbol g}_e^{\rm{H}}({\boldsymbol \rho}[n])}
({\boldsymbol B}^*[n]  + {\boldsymbol A}_s^*[n] - {\bar{\boldsymbol{B}}[n]})) +\sigma^2\big)\\
 \nonumber=&{\log_2}\big({{\rm{tr}}({{\boldsymbol g}_u({{\boldsymbol \rho}[n]})}{{\boldsymbol g}_u^{\rm{H}}({\boldsymbol \rho}[n])}{\boldsymbol B}^*[n]}) \\
 \nonumber+& {\rm{tr}}({{\boldsymbol g}_u({{\boldsymbol \rho}[n]})}{{\boldsymbol g}_u^{\rm{H}}({{\boldsymbol \rho}[n]})}{\boldsymbol A}_s^*[n]) +\sigma^2\big)\\
 + &{\log_2}\big({\rm{tr}}({{\boldsymbol g}_e({\boldsymbol \rho}[n])}{{\boldsymbol g}_e^{\rm{H}}({\boldsymbol \rho}[n])}{\boldsymbol A}_s^*[n]) +\sigma^2\big).
\end{align}

Similarly, for the last two terms of Eq. \eqref{eq:object_function_of_p1}, it can be reformulated as
\begin{align}\label{last_two_terms_object_function}
\nonumber&\big({\delta}_u^{(l)}[n] 
 + {\rm{tr}}({\Lambda}_u^{(l)}[n]({\bar{\boldsymbol A}}_s[n] - {\boldsymbol A}_s^{(l)}[n]))\big)\\
\nonumber+&\big({\delta}_e^{(l)}[n] + {{\rm{tr}}({\Lambda}_e^{(l)}[n]({\bar{\boldsymbol{B}}[n]} - {\boldsymbol B}^{(l)}[n]))} \\
\nonumber +& {\rm{tr}}({\Lambda}_e^{(l)}[n]({\bar{\boldsymbol A}}_s[n] - {\boldsymbol A}_s^{(l)}[n]))\big)\\
 \nonumber=&\big({\delta}_u^{(l)}[n] 
 + {\rm{tr}}({\Lambda}_u^{(l)}[n](({\boldsymbol B}^*[n]  + {\boldsymbol A}_s^*[n] - {\bar{\boldsymbol{B}}[n]})- {\boldsymbol A}_s^{(l)}[n]))\big)\\
\nonumber+&\big({\delta}_e^{(l)}[n] + {{\rm{tr}}({\Lambda}_e^{(l)}[n]({\bar{\boldsymbol{B}}[n]} - {\boldsymbol B}^{(l)}[n]))} \\
\nonumber +& {\rm{tr}}({\Lambda}_e^{(l)}[n](({\boldsymbol B}^*[n]  + {\boldsymbol A}_s^*[n] - {\bar{\boldsymbol{B}}[n]}) - {\boldsymbol A}_s^{(l)}[n]))\big)\\
\nonumber =&\big({\delta}_u^{(l)}[n] 
 + {\rm{tr}}({\Lambda}_u^{(l)}[n]({\boldsymbol A}_s^*[n] - {\boldsymbol A}_s^{(l)}[n]))\big)\\
\nonumber+&\big({\delta}_e^{(l)}[n] + {{\rm{tr}}({\Lambda}_e^{(l)}[n]({\boldsymbol B}^*[n] - {\boldsymbol B}^{(l)}[n]))}\\
+& {\rm{tr}}({\Lambda}_e^{(l)}[n]({\boldsymbol A}_s^*[n] - {\boldsymbol A}_s^{(l)}[n]))\big).
\end{align}

By combining \eqref{first_two_terms_object_function} and \eqref{last_two_terms_object_function}, it is evident that the objective value remains the same. 

This completes the proof.


\begin{thebibliography}{12}
\bibliographystyle{IEEEbib}
\bibitem{LiuIntegrated2022}
F.~Liu, Y.~Cui, C.~Masouros, J.~Xu, T.~Han, Y.~Eldar, and S.~Buzzi, ``Integrated sensing and communications: {Toward} dual-functional wireless networks for {6G} and beyond,'' \emph{IEEE J. Sel. Areas Commun.}, vol.~40, no.~6, pp. 1728-1767, Mar. 2022.

\bibitem{LyuJoint2022}
Z.~Lyu, G.~Zhu, and J.~Xu, ``Joint maneuver and beamforming design for {UAV}-enabled integrated sensing and communication,'' \emph{IEEE Trans. Wireless Commun.}, vol.~22, no.~4, pp. 2424-2440, Oct. 2022.

\bibitem{MengUAV2024}
K.~Meng, Q.~Wu, J.~Xu, W.~Chen, Z.~Feng, and R.~Schober, ``UAV-enabled integrated sensing and communication: Opportunities and challenges,"\emph{IEEE Wireless Commun.}, vol.~31, no.~2, pp.~97-104, Apr.~2024.

\bibitem{KhaliliEfficient2024}
A.~Khalili, A.~Rezaei, D.~Xu, F.~Dressler, and R.~Schober, ``Efficient UAV hovering, resource allocation, and trajectory design for ISAC with limited backhaul capacity,"  \emph{IEEE Trans. Wireless Commun.}, vol.~23, no.~11, pp.~17635-17650, Nov.~2024.

\bibitem{KhaliliEnergy2023}
A.~Khalili, A.~Rezaei, D.~Xu, and R.~Schober, ``Energy-aware resource allocation and trajectory design for UAV-enabled ISAC," in \emph{Proc. IEEE
Global Commun. Conf. (GLOBECOM)}, Dec.~2023, pp.~4193-4198.

\bibitem{ZhongSecure2018}
C.~Zhong, J.~Yao, and J.~Xu, ``Secure {UAV} communication with cooperative
 jamming and trajectory control,'' \emph{IEEE Commun. Lett.}, vol.~23, no.~2, pp. 286-289, Feb. 2019.

\bibitem{WuWhen2023}
J.~Wu, W.~Yuan and L.~Hanzo, ``When {UAVs} meet {ISAC}: Real-time trajectory design for secure communications,'' \emph{IEEE Trans. Veh. Technol.}, vol. 72, no. 12, pp. 16766-16771, Dec. 2023.

\bibitem{RenRobust2023}
Z.~Ren, L.~Qiu, J.~Xu and D.~W.~K.~Ng, ``Robust transmit beamforming for secure integrated sensing and communication,'' \emph{IEEE Trans. Commun.}, vol. 71, no. 9, pp. 5549-5564, Sep. 2023.

\bibitem{LiuSecure2025}
Y.~Liu, X.~Liu, Z.~Liu, Y.~Yu, M.~Jia, and Z. Na, ``Secure rate maximization for {ISAC-UAV} assisted communication amidst multiple eavesdroppers,'' \emph{IEEE Trans. Veh. Technol.}, vol. 73, no. 10, pp. 15843-15847, Oct. 2024.

\bibitem{ZouSecuring2024}
J.~Zou, C.~Masouros, F.~Liu, and S.~Sun, ``Securing the sensing functionality in {ISAC} networks: {An} artificial noise design,'' \emph{IEEE Trans. Veh. Tech.}, vol. 73, no. 11, pp. 17800-17805, Nov. 2024.

\bibitem{RenSecure2024}
Z.~Ren, J.~Xu, L.~Qiu, and D.~W.~K.~Ng, ``Secure cell-free integrated sensing and communication in the presence of information and sensing eavesdroppers,'' \emph{IEEE J. Sel. Areas Commun.}, vol.~42, no.~11, pp.~3217-3231, Nov.~2024.

\bibitem{Mukherjee2012}
A.~Mukherjee and A.~L. Swindlehurst, ``Detecting passive eavesdroppers in the {MIMO} wiretap channel,'' in \emph{Proc. IEEE Int. Conf. Acoustics, Speech, and Signal Process.}, Mar. 2012, pp. 2809-2812.

\bibitem{ShangUnmanned2019}
B.~{Shang}, L.~{Liu}, J.~{Ma}, and P.~{Fan}, ``Unmanned aerial vehicle meets vehicle-to-everything in secure communications,'' \emph{IEEE Commun. Mag.}, vol.~57, no.~10, pp. 98-103, Oct. 2019.

\bibitem{LinThe2018}
X. Lin, V. Yajnanarayana, S. D. Muruganathan, S. Gao, H. Asplund, H. Maattanen, M. Bergstrom, S. Euler, and Y. E. Wang, ``The sky is not the limit: LTE for unmanned aerial vehicles," {\it IEEE Commun. Mag.}, vol. 56, no. 4, pp. 204-210, Apr. 2018.

\bibitem{3GPP}
3GPP TR 36.777, ``Study on enhanced LTE support for aerial vehicles," Dec. 2017.


\bibitem{LiuJoint2020}
X.~Liu, T.~Huang, N.~Shlezinger, Y.~Liu, J.~Zhou and Y.~C.~Eldar, ``Joint transmit beamforming for multiuser MIMO communications and MIMO radar,'' \emph{IEEE Trans. Signal Process.}, vol. 68, pp. 3929-3944, Jun. 2020.
\end{thebibliography}
\end{document}